# Broadband vibrational sum-frequency generation spectrometer at 100 kHz in the 950-1750 cm$^{-1}$ spectral range utilizing a LiGaS$_2$ optical parametric amplifier

**ZSUZSANNA HEINER,**[1,*] **LI WANG,**[2] **VALENTIN PETROV,**[2] **AND MARK MERO**[2]

[1]*School of Analytical Sciences Adlershof, Humboldt-Universität zu Berlin, Albert-Einstein-Str. 5-11, 12489 Berlin, German*
[2] *Max Born Institute for Nonlinear Optics and Short Pulse Spectroscopy, Max-Born-Str. 2, 12489 Berlin, Germany*
*\* heinerzs@hu-berlin.de*

**Abstract:** We present a 100 kHz broadband vibrational sum-frequency generation (VSFG) spectrometer operating in the 5.5-10.5 µm (950-1750 cm$^{-1}$) wavelength range. The mid-infrared beam of the system is obtained from a collinear, type-I LiGaS$_2$ crystal based optical parametric amplifier seeded by a supercontinuum and pumped directly by 180 fs, ~32 µJ, 1.03 µm pulses from an Yb:KGd(WO$_4$)$_2$ laser system. Up to 0.5 µJ mid-infrared pulses with durations below 100 fs were obtained after dispersion compensation utilizing bulk materials. We demonstrate the utility of the spectrometer by recording high-resolution, low-noise vibrational spectra of Langmuir-Blodgett supported lipid monolayers on CaF$_2$. The presented VSFG spectrometer scheme offers superior signal-to-noise ratios and constitutes a high-efficiency, low-cost, easy-to-use alternative to traditional schemes relying on optical parametric amplification followed by difference frequency generation.

## 1. Introduction

Robust, commercially available diode-pumped (sub-)ps Yb lasers have been gaining foothold as the main workhorse behind spectroscopic applications requiring high peak and average power. In particular, down-conversion of the output of Yb lasers to femtosecond mid-infrared (MIR) pulses in the 2.5-12.5 µm (800-4000 cm$^{-1}$) spectral range holds great promise for vibrational spectroscopy, especially broadband vibrational sum-frequency generation (BB-VSFG) and 2D-IR spectroscopy. The superior power scalability of Yb lasers compared to Ti:sapphire lasers has enabled the extension of the repetition rate to 100 kHz in these applications, offering potentially higher signal-to-noise ratios (SNR's) and shorter acquisition times for a wide range of samples. Only a handful of 100 kHz vibrational spectrometers have been demonstrated so far. A home-built BB-VSFG spectrometer, relying on oxide crystal based optical parametric amplifiers (OPA's) pumped at 1 µm, operated in the 2.8-3.6 µm range at a pump-to-MIR energy conversion efficiency of several %, but the long wavelength edge of the spectrum was limited by the absorption edge of the amplifying material to ~4.5 µm [1]. Vibrational spectrometers operating at longer wavelengths have until now relied on optical parametric amplification in oxide nonlinear materials followed by difference frequency generation (DFG) between the signal and idler in a non-oxide material, which limited the pump-to-MIR energy conversion efficiency to ~0.5% [2, 3].

Eliminating the DFG step in generating MIR pulses above 5 µm could potentially increase the conversion efficiency. However, known oxide nonlinear crystals exhibit strong phonon absorption in this spectral range. While most non-oxide crystals that are transparent above 5 µm suffer from strong residual or two-photon absorption when pumped at 1 µm, there are still few commercially available options that can be pumped directly by Yb lasers [4]. LiGaS$_2$ (LGS)

[5, 6] has recently been successfully used in OPA's pumped at 1.03 µm. So far, the demonstrated MIR pulse energies were limited to a few 10 nJ [7, 8]. Furthermore, the shortest measured pulse duration above a center wavelength of 5 µm was ~200 fs [7]. None of the aforementioned systems have been applied for vibrational spectroscopy thus far.

VSFG spectroscopy is a surface-specific nonlinear optical tool, where the interface sensitivity arises from the vanishing $\chi^{(2)}$ response of an isotropic, centrosymmetric bulk medium and the detection and analysis of even sub-monolayers of interfacial molecules is possible [9-11]. The technique applies two incoming electromagnetic fields. One beam is resonant with the molecular vibrations in the MIR range, typically 800-4000 cm$^{-1}$, while the second beam consists of picosecond, visible (VIS) pulses, the bandwidth of which determines the resolution in the measured vibrational spectrum. Here, the term "visible," as is customary in VSFG spectroscopy, is used to denote both visible and near-infrared wavelengths. In a broadband VSFG experiment [12], the narrowband visible and the broadband MIR laser pulses are temporally and spatially overlapping at the surface. Through the SFG process, the generated vibrational fingerprint excited by the femtosecond MIR pulses is upconverted to the visible range, where technologically mature Si-based array detectors operate. For the generation of the narrowband VIS pulses, typically lossy spectral filtering of the broadband pump pulses is employed via bandpass interference filters, etalons, or zero-dispersion monochromators [13]. In our previous work, we demonstrated and applied the first 100 kHz BB-VSFG system, which operated in the O-H, C-H stretching vibrational region between 2.8-3.6 µm yielding unprecedentedly high signal-to-noise ratios and short acquisition times [1, 14]. For generating the VIS pulses, we employed a highly efficient spectral compression technique based on chirped SFG and obtained a bandwidth of only 3 cm$^{-1}$, an order of magnitude below the resolution of standard BB-VSFG spectrometers.

Here, we present a 100 kHz BB-VSFG spectrometer, with an extended MIR spectral coverage of 6-10 µm utilizing a single-stage, LGS-based OPA pumped directly at 1 µm. We demonstrate an order of magnitude increase in pulse energy combined with the first sub-100 fs pulses in the 7-8.5 µm spectral range from an Yb-laser-pumped LGS OPA. Up to 48 mW (i.e., 0.48 µJ) of MIR average power is available for experiments with pulse durations down to 98 fs. We experimentally demonstrate the feasibility of our scheme by presenting high-resolution BB-VSFG spectra of a solid-supported lipid monolayer system in the fingerprint region between 950 and 1350 cm$^{-1}$ without tuning the central MIR frequency, yielding very high SNR's within short acquisition times.

## 2. Methods

### 2.1 Broadband VSFG spectrometer

Our home-built BB-VSFG spectroscopic setup consists of (i) a single-stage OPA system providing ~100 fs pulses at center wavelengths between 6 and 10 µm and (ii) a chirped SFG stage producing narrowband picosecond pulses at 515 nm. The pump source is a commercial, 6 W Yb:KGd(WO$_4$)$_2$ laser (Pharos, Light Conversion, Ltd.) operating at a central wavelength of 1030 nm and a repetition rate of 100 kHz, which drives the entire VSFG spectroscopic system. The schematic of the BB-VSFG spectrometer is shown in Fig. 1(a) and 1(b). The pump pulses are split into two parts using a partial reflector: a 40 µJ portion is applied to pump the single-stage OPA, while the remaining 20 µJ part is used to produce narrow bandwidth, picosecond visible pulses.

For the generation of the picosecond visible pulses from femtosecond near-infrared pulses, we employ an efficient SFG spectral compression scheme [15, 16]. Our implementation of the spectral compression technique is explained in more detail in Ref. [1]. Briefly, the 20 µJ, 180 fs pulses are split into two equal parts and are sent into a positive and a negative dispersion unit (i.e., with positive and negative group delay dispersion, respectively). After chirp manipulation, the beams are focused into a 1.5 mm-thick, type-I BBO crystal, producing 5 µJ, transform-limited, 4.5 ps pulses at a center wavelength of ~515 nm. We achieved a high overall conversion efficiency of 25%, while the spectral width is reduced from the original value of ~80 cm$^{-1}$ to below 3 cm$^{-1}$ which is a highly cost-efficient way to produce very narrowband visible pulses. In contrast, VIS pulse bandwidths in broadband VSFG spectrometers are usually 15-30 cm$^{-1}$, achieved at high filtering losses [13].

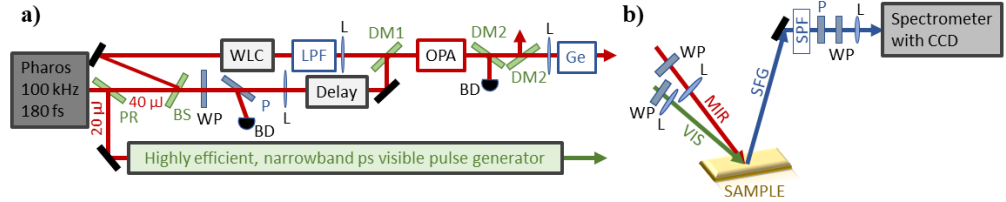

Fig. 1. Schematic layout of the BB-VSFG spectrometer. (a) Broadband MIR and narrowband visible pulse generation units. (b) VSFG unit. BS: beam sampler, PR: partial reflector, WP: half-wave plate, P: polarizing beam splitter, WLC: white light continuum generation unit, LPF: long-pass filter, SPF: short-pass filter, L: lens, BD: beam dump. DM1: dichroic mirror, high reflection (HR) at 1.03 µm and high transmission (HT) at >1.1 µm, DM2: dichroic ZnSe mirror, HR at 1.0-1.2 µm, and HT at 6-12 µm, Ge: germanium-based pulse compression unit. All lenses, wave plates, and filters are AR-coated.

For OPA seed generation, a small fraction of the 40 µJ pulses is focused into a 6 mm-thick uncoated YAG plate to generate a supercontinuum on the Stokes-side of the pump pulses resulting in optically synchronized seed pulses at the signal wavelength [17]. After collimation with a parabolic mirror (not shown in the figure), a long-pass filter is used to block the pump beam and transmit the long wavelength part of the continuum (i.e., >1.1 µm). The single-stage OPA is based on a 5 mm-long, uncoated, type-I LGS crystal (ooe phase-matching in x-z principal plane, $\theta = 48.2°$). The supercontinuum pulses are focused by an $f = 500$ mm lens through a dichroic mirror (cf. DM1 in Fig. 1(a)), while the 31.8 µJ pump pulses are focused by an $f = 1000$ mm lens into the LGS crystal. The energy content of the supercontinuum in the 1150-1200 nm range incident on the OPA crystal was measured to be ~13 nJ. The beam waist radii (1/e² values) of the seed and pump pulses at the crystal are 480 and 500 µm, respectively. Despite the high pump intensity (i.e., 45 GW/cm$^2$ on-axis peak value), we did not observe crystal damage. An output idler pulse energy of 0.53 µJ (i.e., 53 mW average power) at 7.8 µm was measured after two dichroic mirrors (cf. DM2 in Fig. 1(a)), an $f = 200$ mm germanium re-collimating lens, and an uncoated, 1 mm-thick germanium window used at the Brewster-angle (not shown in Fig. 1(a)). The pump-to-signal energy conversion efficiency, when corrected for the losses introduced by the uncoated LGS surfaces and the two dichroic mirrors is calculated to be ~2.5%, corresponding to a quantum conversion efficiency of ~19%. The OPA produces broad bandwidth MIR laser pulses with a center wavelength tunable from 6 to 10 µm (covering a full frequency range of 950-1750 cm$^{-1}$). Figure 2(a) shows the spectral tunability of our MIR pulses. The MIR spectra were measured with an optical spectrum analyzer (OSA207C, Thorlabs, Inc.) directly after the germanium Brewster window. The full width at the half maxima are in the range of 120-200 cm$^{-1}$. The spectral tuning of the MIR pulses is achieved by turning the LGS crystal and optimizing the spatial and temporal overlap between the seed and pump pulses. Figure 2(b) shows the average power of the chirp-compensated MIR pulses as a function of center wavelength. The MIR pulse energies after the pulse compression unit, that is based on AR-coated Ge plates, are measured to be 0.48 µJ (i.e., 48 mW) in the spectral region of 7.8 µm and decrease to 0.1 µJ (i.e., 10 mW) at center wavelengths of around 6.5 and 8.7 µm. We attribute the substantial drop in conversion efficiency at the edges of our tuning range mainly to two factors. First, the increasing group velocity mismatch between pump and signal/idler pulses when the idler wavelength is tuned away from 7.8 µm [5] leads to a decrease in the effective crystal length. Second, the transmission window of the setup is limited by DM2 on the short wavelength side and by the onset of absorption of LGS on the long wavelength side [18].

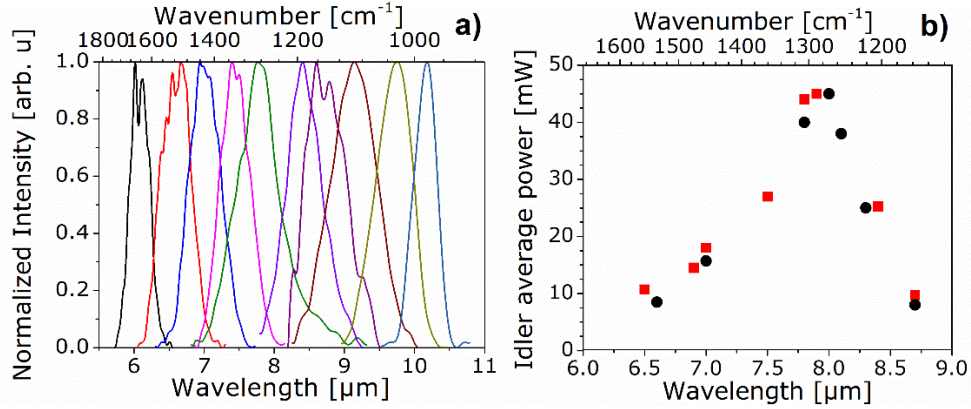

Fig. 2. (a) Spectral tunability of the broadband MIR pulses obtained with an optical spectrum analyzer. (b) Chirp-compensated idler average power as a function of center wavelength.

The duration of the MIR pulses at different central wavelengths was measured by a home-built cross-correlation frequency-resolved optical gating (X-FROG) apparatus. The X-FROG measurements were performed using noncollinear sum-frequency generation in a 200 µm-thick, type-I LGS (ooe phase-matching in x-z principal plane, $\theta = 43°$) crystal. Residual 180 fs, 1.03 µm pulses from the Yb:KGW pump laser, characterized by a home-built second-harmonic FROG device, served as the gate pulses. Since the generated MIR idler pulses are negatively chirped, the compression is easily feasible using Ge windows with a total length of 11 mm. The measured pulse width was 98 fs corresponding to 3.8 optical cycles at 7.8 µm and increased to 109 and 112 fs at a central wavelength of 7.2 and 8.2 µm, respectively. The measured and retrieved FROG traces together with the measured spectral and reconstructed temporal and spectral profiles of the amplified idler pulses at 7.2, 7.8, and 8.2 µm are shown in Fig. 3. In addition, we show in the middle column also the chirp-free idler temporal profiles calculated from the directly measured pulse spectra. The blue side of the measured MIR spectra shows signs of atmospheric water absorption starting below ~7.5 µm. The difference between the measured and X-FROG reconstructed spectra is due to the different atmospheric propagation distances in the two cases. The near-field beam profile (shown in Fig. 3, middle panel, inset) of the 7.8 µm, 0.5 µJ laser pulses at a distance of 1.1 m behind the LGS crystal, a collimating lens and the Ge-based compression unit was measured using a pyroelectric camera with a pixel size of 80 µm. The $1/e^2$ radii in the horizontal and vertical planes were 3.7 and 3.0 mm, respectively. The MIR beam is nearly Gaussian with a slight elliptical extension in the horizontal direction determined by the shape of the pump beam profile at the crystal [19]. The area on the optical table used for the single stage OPA source was only 0.15 m².

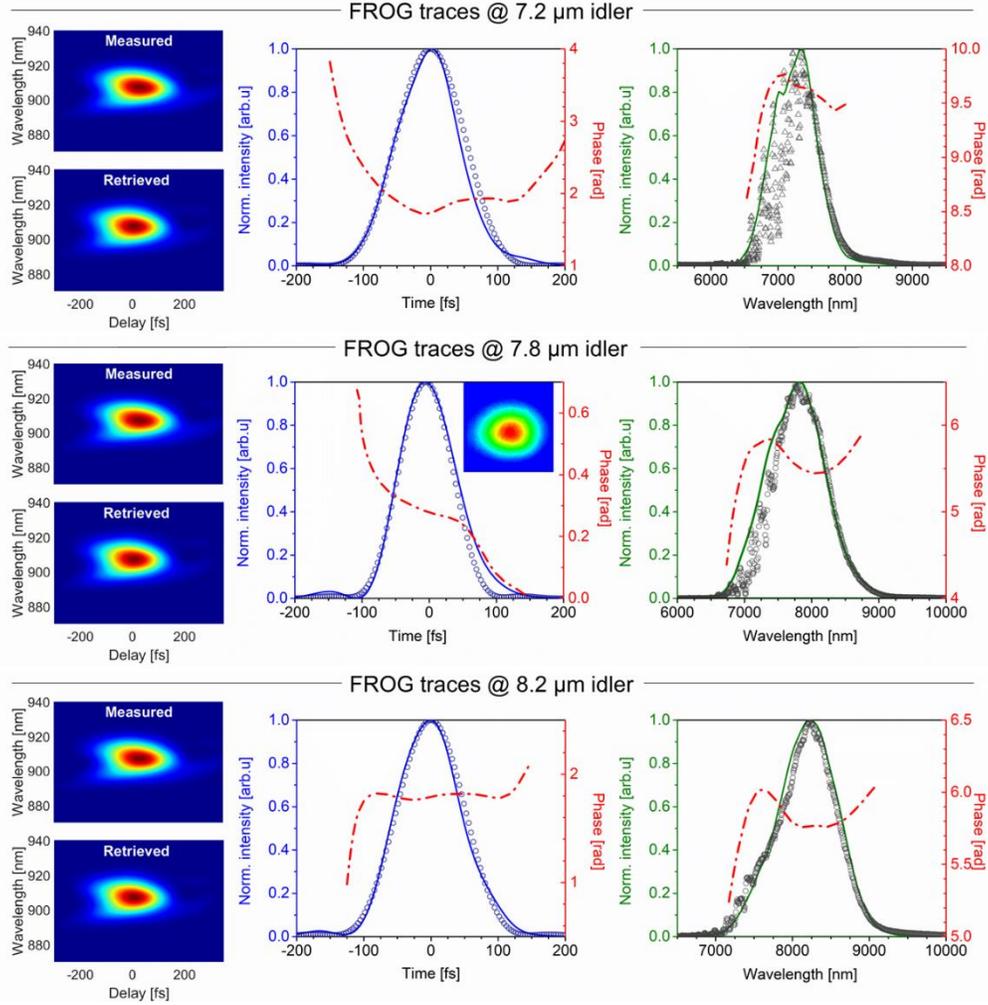

Fig. 3. X-FROG data obtained for the compressed, 109, 98, and 112 fs idler pulses at 7.2, 7.8, and 8.2 µm, respectively. Measured and retrieved (left column) X-FROG spectrograms, and reconstructed temporal (middle column) and spectral (right column) pulse intensity and phase. The directly measured spectrum is shown by the symbols (right column, grey circles), and the symbols in the temporal intensity curves (middle column, blue circles) are the corresponding Fourier transform. The FROG-error of the reconstruction for a grid size of 256 × 256 points was 0.0017, 0.0017, and 0.0014 at 7.2, 7.8, and 8.2 µm, respectively. The inset in the middle graph shows the Gaussian near-field spatial beam profile of the 0.5 µJ pulses at 7.8 µm with a $1/e^2$ diameter of 7.4 and 6.0 mm in the horizontal and vertical plane (i.e., parallel to the y principal axis of LGS), respectively.

## 2.2 BB-VSFG measurements

In our BB-VSFG setup [Fig. 1(b)], the ~100 fs MIR pulses and the 4.5 ps VIS pulses were focused onto the sample surface using an $f = 40$ mm and an $f = 300$ mm lens, respectively. The pulse energy and spot sizes at the sample ($1/e^2$ diameters) were estimated to be ~0.25 µJ and ~28 µm for the MIR beam at 8.0 µm and ~4.15 µJ and ~150 µm for the VIS beam, respectively. The angle of incidence was 57° and 68° relative to the surface normal for the MIR and the VIS beam, respectively. We used zero-order half-wave plates to control the polarizations of the excitation pulses, and a polarizer and a half-wave plate for selecting the polarization component of the generated SFG signal relative to the plane of incidence on the sample (P: parallel, S:

perpendicular) and to provide high diffraction efficiency in the grating spectrograph (see Fig. 1(b)). The SFG spectra were measured in SSP and PPP polarization combinations (i.e., polarizations of the SFG, VIS, and MIR beams, respectively) and recorded with a 320 mm imaging spectrometer equipped with a Peltier-cooled, back-illuminated, deep-depletion CCD (Horiba, Synapse). In all measurements, the so-called 'window-geometry' (cf. Fig. 1(b)) was applied at an air-solid interface [20, 21]. For the normalization of the so-obtained raw VSFG spectrum, a reference spectrum was measured using a 1 mm-thick GaAs plate. For frequency calibration, a polystyrene film with a well-known absorption spectrum was inserted in the MIR beam [22].

The Langmuir-Blodgett (LB) technique was used to prepare 1,2-dipalmitoyl-*sn*-glycero-3-phosphocholine (DPPC) monolayers on a $CaF_2$ surface [23, 24]. The details of the preparation of the LB supported lipid monolayers have been provided elsewhere [25]. In brief, a 1 mg/ml stock solution of DPPC (grade >99%, Avanti Polar Lipids, Inc.) was prepared by dissolving the powder in chloroform:methanol (9:1, v:v). Then a 10 µl stock solution was spread on the water surface using a Hamilton syringe, and a time period of 15 minutes was permitted for the complete evaporation of the solvent before compression was started. A surface pressure of 30 mN/m was applied during the preparation of the monolayer on a $CaF_2$ plate resulting in a liquid condensed phase for the lipid monolayer. A freshly cleaned $CaF_2$ window was dip-coated using the LB method to prepare the monolayer of DPPC on the solid surface at a raising speed of 1 mm/min, while the surface pressure was held constant. The VSFG measurements were performed directly after the preparation process.

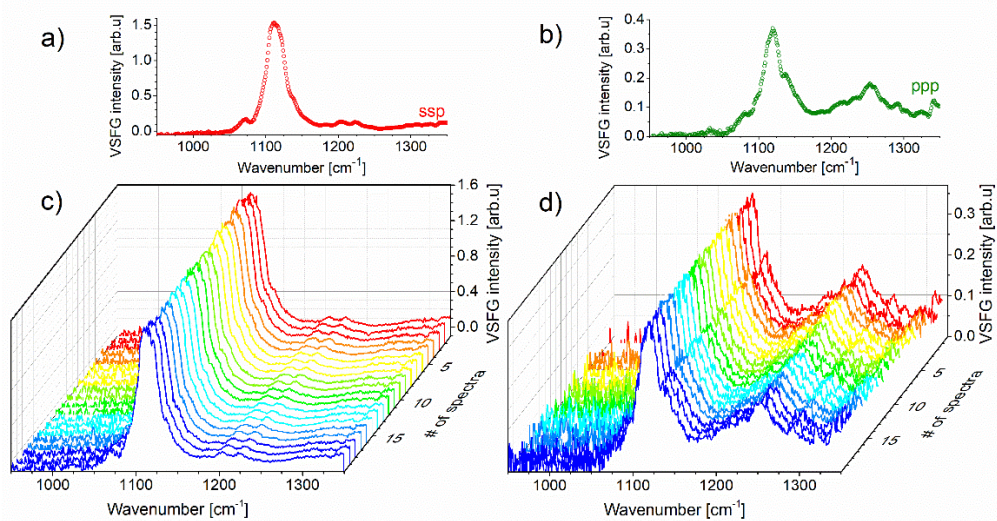

Fig. 4. Normalized VSFG spectra of a DPPC monolayer at the air-solid interface using (a,c) SSP and (b,d) PPP polarization combinations at a pulse repetition rate of 100 kHz obtained at a central MIR frequency of 1246 cm$^{-1}$ (i.e., 8.0 µm). The acquired 20 single spectra are shown in (c,d), and the averages of them are displayed in (a,b). The acquisition time was 30 seconds per spectrum. The signal-to-noise ratio is 262 and 70 in (a) and (b), respectively.

To demonstrate the applicability of our 100 kHz spectrometer, the BB-VSFG spectra of the DPPC monolayer at the air-$CaF_2$ interface were collected in the spectral range between 950 and 1350 cm$^{-1}$. Figure 4(a) and 4(b) show the average of twenty VSFG spectra, and Fig. 4(c) and 4(d) display twenty representative single spectra (acquired at a 30 second integration time) obtained at a MIR central frequency of 1246 cm$^{-1}$ in SSP and PPP polarization combinations. In this spectral region, the symmetric and asymmetric $PO_2^-$ stretching modes dominate at 1100 cm$^{-1}$ and 1250 cm$^{-1}$, respectively. While the asymmetric stretching vibrational mode around

1250 cm$^{-1}$ interferes constructively in PPP polarization combination, this mode shows destructive interference, i.e., a negative amplitude, in SSP polarization combination. A similar interference effect was observed on 2-distearoyl-*sn*-glycero-phosphatidylcholin (DSPC) and 1,2-distearoyl-*sn*-glycero-phosphatidylserine (DSPS) monolayers earlier [26]. Under our experimental conditions, the surface coverage of DPPC molecules was ~40 Å$^2$/molecule at an applied surface pressure of 30 mN/m. The average powers of the VIS and MIR beams incident on the sample were 415 mW and 25 mW, respectively. When applying an acquisition time of 30 seconds for a single vibrational spectrum, we achieved SNR's of 69 for SSP and 22 for PPP polarization configurations. Here, the SNR was calculated as the ratio of the amplitude of the symmetric PO$_2^-$ stretching band at 1104 cm$^{-1}$ and the RMS noise floor. The RMS noise was estimated based on the spectral range between 950 and 1000 cm$^{-1}$. When averaging 20 single vibrational spectra, the SNR increased to 262 and 70 in SSP and PPP polarization configurations, respectively. The acquisition time of 20 single vibrational spectra was only 10 minutes, which is significantly shorter than what is possible using 1 kHz BB-VSFG or scanning VSFG systems when attempting to approach our SNR's [27, 28].

Obtaining vibrational spectra at high spectral resolution and SNR is especially important when trying to retrieve information on the structure of the interfacial molecular layer and the orientation of functional molecular groups relative to the surface normal. In the molecular fingerprint region, where vibrational spectra are more congested than in the C-H stretching region, the requirement of high-quality vibrational spectra is more stringent. The spectral resolution and SNR provided by our BB-VSFG spectrometer compare favorably with those possible with standard VSFG systems [27, 29]. Our measurements resolved the presence of eleven vibrational bands of DPPC in the 950-1350 cm$^{-1}$ region, a higher number than previously observed. Thus, using high laser repetition rate and high spectral resolution enabled by the spectral compression scheme in the VIS beam line hold great promise for a more detailed investigation of a wide range of monolayers at small surface coverage not only in the C-H, O-H stretching region [14, 25], but also in the molecular fingerprint region.

## 3. Conclusions

We have presented a high-resolution BB-VSFG spectrometer operating at a laser repetition rate of 100 kHz in the spectral fingerprint region between 950 and 1750 cm$^{-1}$ for the first time. The spectrometer pumped by only 60 µJ at 1.03 µm consists of (i) a broadband, ~100 fs, tunable MIR beam, and (ii) a 3 cm$^{-1}$-bandwidth, picosecond VIS beam centered around 515 nm. The MIR pulses were generated in a single-stage, LGS-based OPA producing 0.1-0.5 µJ idler energies in the 6.5-8.75 µm wavelength range, at a maximum quantum conversion efficiency of 19% at 7.8 µm in a very compact and simple setup. These pulse energies are based on actual measurements performed at the output of the spectrometer and are available for VSFG spectroscopy and other applications. We demonstrated sub-100 fs pulses at 7.8 µm corresponding to 3.8 cycles, which constitute the shortest MIR pulses obtained from an Yb-laser-pumped LGS amplifier. By employing direct OPA pumping at 1 µm without utilizing an additional DFG stage, the overall conversion efficiency was increased beyond that of standard schemes relying on OPA followed by DFG. In contrast to the aforementioned systems, the present MIR pulse generation scheme also offers passive carrier-envelope phase stability, which, together with the few-cycle nature of the idler pulses can be exploited in strong-field experiments on condensed phase systems, such as high-harmonic generation in solids. The high stability and the two orders of magnitude higher laser repetition rate compared to typical available BB-VSFG setups, i.e., Ti:sapphire-laser-based systems, significantly improved the SNR in VSFG spectroscopic applications, while the measurement time decreased by at least a factor of 10. Although the applied total pump pulse energy was limited, the average powers of both the MIR and the VIS pulses reached 50 and 500 mW, respectively, that can drive a broad range of nonlinear and/or vibrational spectroscopic applications. We demonstrated, that our tunable, high-resolution and high-repetition-rate scheme has both better spectral resolution and higher SNR compared to traditional BB-VSFG spectrometers. Applying the presented BB-VSFG spectrometer layout can open up new perspectives in surface/interface sciences, especially in heterogeneous catalysis and on organic interfaces.

## Funding

Funding by the Deutsche Forschungsgemeinschaft (DFG), No. GSC 1013 SALSA and No. PE 607/14-1 is gratefully acknowledged. Z.H. acknowledges funding by a Julia Lermontova Fellowship from DFG, No. GSC 1013 SALSA.

## Acknowledgments

The Authors thank Freeda Yesudas for preparing the LB supported monolayers.